# A field study of data analysis exercises in a bachelor physics course using the internet platform VISPA

**Martin Erdmann [1], Robert Fischer [1], Christian Glaser [1], Dennis Klingebiel [1], Raphael Krause [1], Daniel Kuempel [1], Gero Müller [1], Marcel Rieger [1], Jan Steggemann [1,2], Martin Urban [1], David Walz [1], Klaus Weidenhaupt [1], Tobias Winchen [1] and Birgitta Weltermann [3]**

[1] Physikalisches Institut 3A, RWTH Aachen University, Aachen, Germany

[2] now at CERN, European Organization for Nuclear Research, Geneva, Switzerland

[3] Institute for General Medicine, University of Duisburg-Essen, Essen, Germany

E-mail: erdmann@physik.rwth-aachen.de

**Abstract.** Bachelor physics lectures on "Particle Physics and Astrophysics" were complemented by exercises related to data analysis and data interpretation at the RWTH Aachen University recently. The students performed these exercises using the internet platform VISPA, which provides a development environment for physics data analyses. We describe the platform and its application within the physics course, and present the results of a student survey. The students' acceptance of the learning project was positive. The level of acceptance was related to their individual preference for learning with a computer. Furthermore, students with good programming skills favor working individually, while students who attribute themselves having low programming abilities favor working in teams. The students appreciated approaching actual research through the data analysis tasks.

**Keywords:** university, teaching, online, physics education, internet-based education, cooperative learning

**PACS:** 01.40.-d, 01.40.Di, 01.40.gb, 01.50.H-, 01.50.hv

# 1. Introduction

Today's physics students typically grew up with smartphones, computers and the internet for acquisition of knowledge. Also, many are familiar with the use of virtual and social networking systems. The application of computing skills for learning has already begun and holds considerable potential for educational processes and creative scientific developments.

Therefore, modern teaching concepts combine the proven training methods - lectures, exercises, presentations - with computer-based information channels (Wen-Yu Lee *et al* 2011). The so-called blended learning includes "learning, communication, information, and knowledge management independent of location and time in combination with exchange of experiences,… and personal meetings in classic face to face training situations" (Bender *et al* 2002).

For undergraduate students in physics, blended learning can be suited in advanced physics courses by including analysis of experimental data and their interpretation. The students can explore actual research questions, and experience physics concepts through a different learning channel complementing lectures, books, and classical homework. At this advanced stage, the students are familiar with data analyses, e.g. through previous laboratory courses.

An example of a learning objective is Einstein's relation $E=mc^2$. A deeper understanding of this relation can be achieved by analyzing experimental data of electron-positron pairs revealing J/Psi and Z particles in proton-proton collisions at the Large-Hadron-Collider at CERN.

Interestingly, such questions are already discussed within so-called Masterclasses for high school students at the appropriate level of comprehension (Johansson *et al* 2007). These classes are organized by CERN on a yearly basis. They already use internet platforms enabling interactive visualizations of particle collisions in the particle detectors at CERN, and include studies of certain pre-defined distributions. The high school students get invited for a day to a neighboring university, and are provided with introductory lectures and with computers.

Advanced discussions and analysis applications of physics questions of the type mentioned above usually happen at universities in the final phase of the bachelor level or beyond. Corresponding courses take place in dedicated computer rooms with special software installations suited to programmable data analysis. The number of participants is usually limited by the number of computers provided locally.

With the new VISPA internet platform, (Bretz *et al* 2012a) and (Bretz *et al* 2012b), the number of participants in programmable data analyses is unlimited. Students can use their own internet devices, or alternatively computer pools provided for students. No software installations are required for the user, while maintenance work relates to the central server system only.

During the winter term 2012/13 the VISPA platform has been used by over hundred $3^{rd}$ year students of the experimental physics course "Particle Physics and Astrophysics" at the RWTH Aachen University (Busse 2012). We report here on the internet platform VISPA, and the new course outline and its data analysis exercises. In a student survey we studied the quality and acceptance of this learning project.

**2. Technical environment**

*2.1 Key features of the internet platform*

The internet platform VISPA is a technical innovation for education in physics data analysis. The acronym stands for "visual physics analysis". The platform provides access to a development environment, and is characterized by several features.

VISPA provides online access to individually programmable physics data analyses using web browsers. It supports the entire work flow of a data analysis, including modular design of the program structure, program execution, and reviewing the results. It can be used with keyboard and mouse of notebooks as well as with tablet computers.

Students can share their analyses on the VISPA platform both among themselves and with faculty members. They can use the VISPA platform for team work in face-to-face meetings, or collaborate virtually regardless of location and time.

*2.2. Data analysis environment*

Physicists typically develop their data analysis in several cycles as the first attempt usually does not produce the complete results. These steps include:

1. Draft of the analysis strategy and program logic.

2. Programming and configuration of each module forming a part of the program logic.

3. Running the data analysis program.

4. Verification of the results.

After one cycle, physicists usually realize that corrections or optimizations from step 1, or step 2 are required.

Such development cycles are visually supported by the VISPA platform. Technically, a server-client system has been implemented that uses the web browser of the user as a client. Files containing data, program code, or resulting distributions are stored in the file system of the server. They can be inspected and downloaded (figure 1a).

In the so-called analysis designer, the modular structure of a data analysis is presented as a flow chart (figure 1b). The user drags a module from the list of available modules, drops the module in the working window and connects it to other modules. This resulting picture then provides a visualization of the analysis logic that helps users to keep an overview of their work. It can also serve as a basis for discussion among colleagues. Programming of the individual modules is done in an editor window using the programming language Python (Python Programming Language 2014) (figure 1c).

The data analysis can be executed directly from the analysis designer page. The execution is performed on the server side. The status of the execution can be viewed in a summary table (job dashboard). The visual presentation of data contained in an input file, and data after processing, enables physicists to quickly view details about the data records (data browser window figure 1d). Links to documentation and tutorials can also be selected in the browser window.

With these features the VISPA internet platform covers all functions needed for repetitive iteration of the above-mentioned development cycle of a physics data analysis until the results are satisfactory.

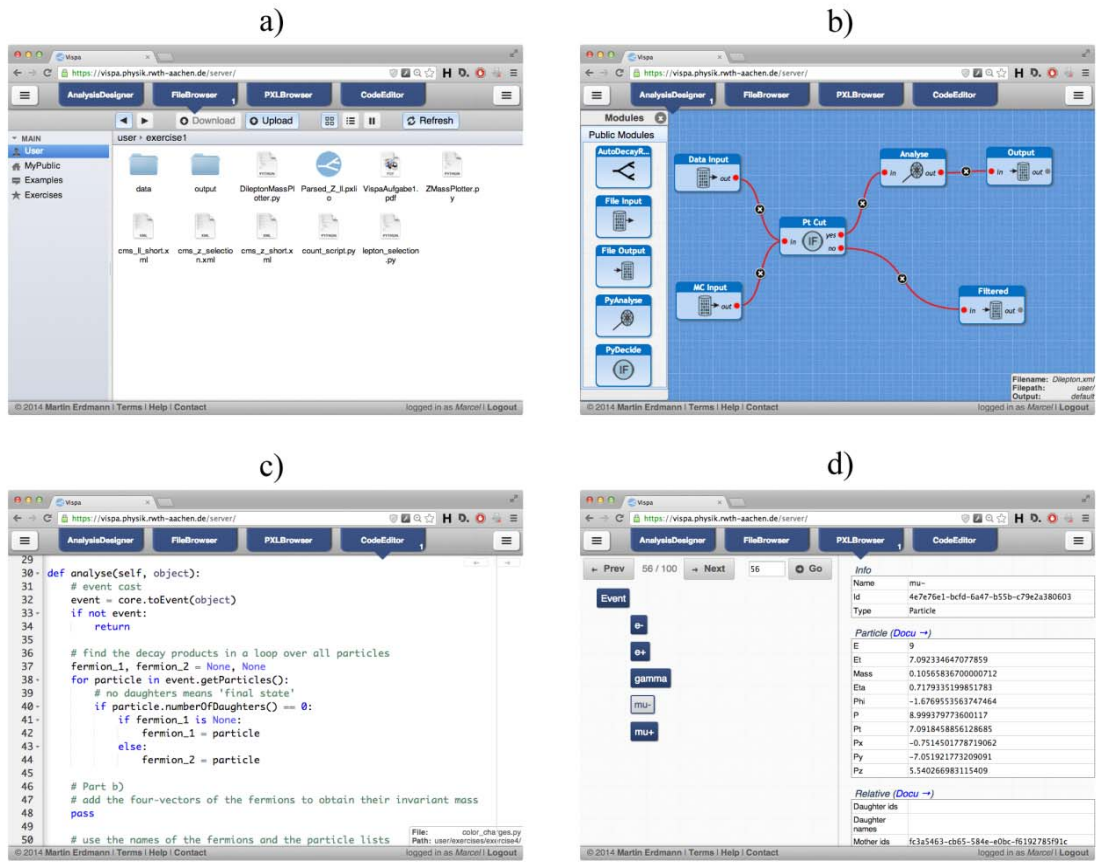

Figure 1: VISPA internet platform for data analysis. a) Window for folders and files, b) analysis designer window with an example module chain, c) text editor for the modules, d) window for inspection of the data.

*2.3. Software libraries*

To make handling of data as comfortable as possible, appropriate interfaces are important prerequisites for creative data analyses. For analyzing a simple laboratory experiment a multicolumn text file is frequently sufficient. For collisions in particle physics experiments, however, the particles need to be described with their characteristics such as momenta, charge, mother-daughter relationships etc., which is stored for each collision event.

The software library PXL (physics extension library) was developed to serve users in handling such data (Bretz *et al* 2012a). It provides appropriate interfaces in the form of C++ classes to perform analyses both with data from simple laboratory experiments as well as data of large experiments in particle and astroparticle physics.

To meet the specific requirements of each experiment and each individual user, a so-called "User Record" was introduced to store objects such as strings, numbers, arrays, matrices etc. under a name chosen by the user. These records can be created and retrieved anytime during the analysis. PXL already offers a number of algorithms for data analysis, but the library is primarily intended to ensure that physicists can develop their own algorithms and ideas for data analysis.

Functionalities presenting frequency distributions are needed for data analyses beyond appropriate representations of the data. A popular toolkit for this purpose is ROOT (Brun and Rademakers 1996) which is used in conjunction with the PXL library. Both libraries have interfaces to the Python language enabling beginners to work with this easy-to-learn programming language.

To support physicists in modular programming, the PXL library also provides a system for control and steering of programmable modules. Modules for various tasks such as reading of particle collision events from a file, or reconstructing a mother particle from its daughter particles can be connected in a single program flow. When the program is executed by the user, the collision data are accessed by a read-in module, and are passed on for further processing from module to module.

The server-client system of VISPA is based on software technologies, which are typically found in web 2.0 applications. All interactions initiated by the students take place on the client side using their favorite web browser (Firefox, Safari, Chrome,…). These browsers support the necessary software JavaScript and HTML 5. For details of the implementation refer to (Bretz *et al* 2012b).

*2.4. Hardware installation*

In order to cope with potentially simultaneous requests of more than 100 students, 10 Linux computers were connected in a network. One of the computers is visible from the internet as the VISPA server. From here, requested tasks are distributed to the other computers within this cluster. All procedures on the system involve a security concept that was created in collaboration with the computing center of the RWTH Aachen University.

The computing power turned out to be sufficient to cope with the type of exercises requested from the students. The CPU usage of the system was even at peak times, i.e. in the evening before delivery of the exercises, well below 10% of the total capacity.

**3. Physics course**

The course on "Particle Physics and Astrophysics" is part of the fifth semester of the bachelor program in physics with 3 hours of lectures per week and exercises due every 2 weeks. To deepen the understanding of the learning material one quarter of the exercises was designed as a data analysis task.

In order to be accepted for the written exam of the course, students had to achieve at least 50% of the exercise points. This value could be reached by exclusively working on the text-based exercises (75% of the achievable points), or alternatively by working on a combination of text-based and data analysis exercises.

*3.1. Practical aspects*

In the beginning of the physics course, the students were introduced to the functionality and applications of the VISPA internet platform in a 90-minute tutorial. This oral presentation included typical examples of programming code using the Python language, e.g. for creating data distributions.

When accessing the VISPA platform for the first time, students were directed to a registration page. After registration, an email was automatically sent to the given email address. The account was opened by the student responding to this email. Each student had a personal folder to develop solutions of the exercises, and a location for the delivery of the exercises. In addition, each student received a folder to share analyses with others for team work.

For each exercise, the students accessed a folder providing the experimental data, and an appropriate example program to start developing their analyses. The students reported the results of the data analysis exercises in the form of distributions and numbers on the written submission together with the text based exercises. In addition, the final program and its results were deposited in the student's delivery folder. As the students frequently work in teams of two, they also had the option of submitting only one common result of their team.

*3.2. Learning objectives*

With the data analyses exercises we extended the physics course and went beyond the capabilities of pen and paper based exercises. The aim was to stimulate additional thinking about the physics concepts presented during the course using the student's familiar computer environment. In addition to the learning objective of Einstein's equation mentioned in the introduction, we included the following topics:

- Understanding the emission of Cherenkov light by charged particles through identifying particles traversing Cherenkov threshold detectors.

- Experiencing time dilation of special relativity by visualizing particles with picosecond life time in the decays of heavy particles.

- Studying the quark fractional electric charges by comparing the energy dependence of hadron production and muon pair production in electron-positron collision data.

- Understanding astronomical observation systems by transforming star positions between horizon, equatorial, and galactic coordinate systems.

- Developing a vision of calorimetric showers in the atmosphere caused by cosmic rays by studying the energy dependence of shower characteristics.

- Approaching fundamental parameters of cosmology by determining the Hubble constant, and the temperature of the cosmic background radiation.

Moreover, the computer-based methods of such data analyses directly relate to current research and provide students with concrete examples about possible directions in their own future research.

*3.3. Example exercise*

To illustrate tasks given to the students, we present the exercise of the aforementioned learning objective of Einstein's energy-mass relationship. The students received a file with 200 collision events of the CMS experiment at CERN containing pairs of electrons or muons. Furthermore, they received an initial data analysis program with examples of accessing properties of the electron pairs in each collision event. This initial program also contained an example how to calculate the so-called invariant mass of the electron pairs. In addition, it presented the tools required to plot a frequency distribution, and finally how to subject the mass distribution to a functional adaptation.

The first task was to describe the resulting distribution in words. Then the range of a Gaussian function had to be adjusted for matching the function to the data distribution. With the function adapted to the data, the mass of the so-called "Z" particle decaying to electron pairs had to be determined from the measured data (example solution figure 2a). The subsequent task was to modify the interval of the invariant mass to [0, 10] GeV/$c^2$. By increasing the number of intervals of the frequency distribution, the mass of another particle, the so-called "J/Psi", became visible. Also the J/Psi mass had to be determined by adapting the function to the data correspondingly (example solution figure 2b).

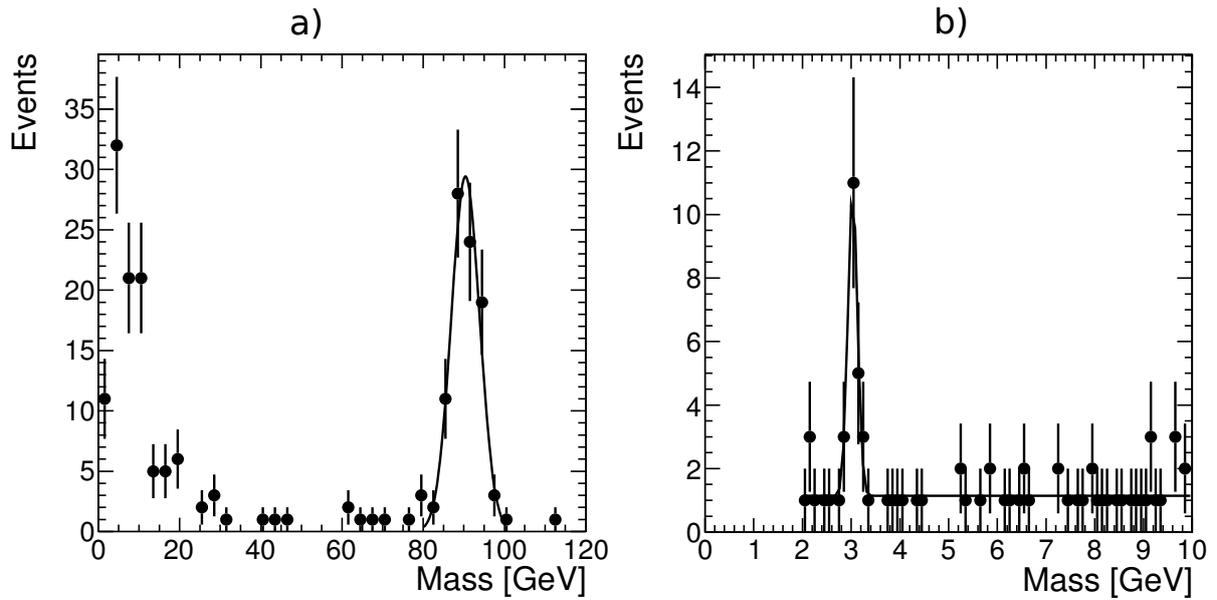

Figure 2: Invariant mass distributions of electron and muon pairs with the CMS experiment at CERN exhibiting a) the Z particle, and b) the J/psi particle from proton-proton collisions.

**4. Survey on the data analysis exercises**

We performed the regular evaluation campaign using the standard questionnaire of the RWTH Aachen University in the middle of the winter term (December 2012). We extended the questionnaire by questions to evaluate the learning project using VISPA. Categories of the questions were related to a) the individual learning style, b) individual versus team work, c) experience with the VISPA technology, d) learning benefit, and e) daily usage of computers and computer programs. The poll was done within a five day time period using the web access of the RWTH evaluation system.

Our main objectives were to learn about the students' usage and acceptance of the VISPA system as well as their experience with the data analysis exercises. We wanted to exploit criteria why students liked or disliked the project, and learn about their personal path of dealing with the new learning approach.

Most questions request an answer on a Likert-scale with 5 options "strongly disagree", "disagree", "undecided", "agree", "strongly agree", which we consider to be an interval scale and present here numerically between -2 and 2. We report in table 1 the fraction of positive answers (options "1" and "2"), and show distributions of the answers where they give additional important information. Moreover, the students were asked to give an overall mark for VISPA using the German grading system, which has been translated here onto a scale between 1 and 5 with the highest mark being the best one. Furthermore, the working hours per exercise and frequencies of accessing the web system per exercise were polled. For each question we provided 5 numerical options. In addition, the students had the possibility to provide free text comments.

## 5. Results of the student survey

In this section, we first report on the participants contributing to the evaluation, and on a self-assessment of their daily usage of computers (table 1). We then describe the evaluation results of the learning project with respect to the techniques of the VISPA internet platform, the students' experience with the exercises, and their evaluation concerning learning style and learning benefit. Finally, we study factors influencing the students' acceptance of the learning project.

### 5.1. Population

The participants of the course were mostly students in their $3^{rd}$ year of bachelor in physics. Out of the 188 registered participants, 135 participated in the computer exercises. 63 students answered the questionnaire (34%), and 61 of them had used the VISPA system for data analysis exercises. The majority of the students are male. About 13% of the students are female. Most of the students are German.

Table 1: Results of the students' survey, more details are given in the text.

|  | **Absolute Numbers** | **Fraction of total %** | |
|---|---|---|---|
| Registered course participants | 188 | 100 | |
| Participated in exercises (received at least 1 point) | 144 | 77 | |
| Participated in computer exercises | 135 | 72 | |
| Participated in questionnaire | 63 | 34 | |
| 3rd year students |  | 87 | |
| Gender male/female |  | 87 | 13 |
| Nationality German/non-German |  | 97 | 3 |
| **Daily usage of computers & computer programs** | **Positive answers (option 1 or 2)** | **Fraction positive %** | |
| Expects data analyses in future | 45 | 75 | |
| Participates in social networks | 39 | 66 | |
| Able to learn at computers well | 36 | 60 | |
| Uses text & calculation programs | 34 | 57 | |
| Has substantial programming experience | 25 | 41 | |
| Likes to play online games | 14 | 24 | |
| **Experience with VISPA technology** |  |  | |
| VISPA server was online | 45 | 76 | |
| Worked in my web browser | 45 | 75 | |
| Satisfied with support | 43 | 75 | |
| Has sufficient programming experience | 34 | 56 | |
| VISPA tutorial was helpful | 29 | 50 | |
| VISPA suited for data analyses | 29 | 48 | |
| Intuitive user interface | 25 | 42 | |
| Sufficient documentation | 20 | 34 | |
| **Individual versus team work** |  |  | |
| Worked on exercises | 46 | 75 | |
| Solved exercises individually | 31 | 52 | |
| Solved in team at same location | 27 | 44 | |
| Solved with virtual team | 7 | 11 | |
| **Learning benefit** |  |  | |
| Provides insight into actual research | 40 | 67 | |
| Deepens content of lectures | 29 | 48 | |
| **Learning style** |  |  | |
| Enjoyed to perform exercises | 28 | 47 | |
| Matches my learning style | 15 | 25 | |

*5.2. Daily usage of computers and computer programs*

Many students reported to use computers in various contexts, e.g. for their social networks (66% positive answers, i.e. answer options 1 or 2). 60% of the students reported to be able to learn well using computers (figure 3a). 41% of the students reported having substantial programming experience (figure 3b).

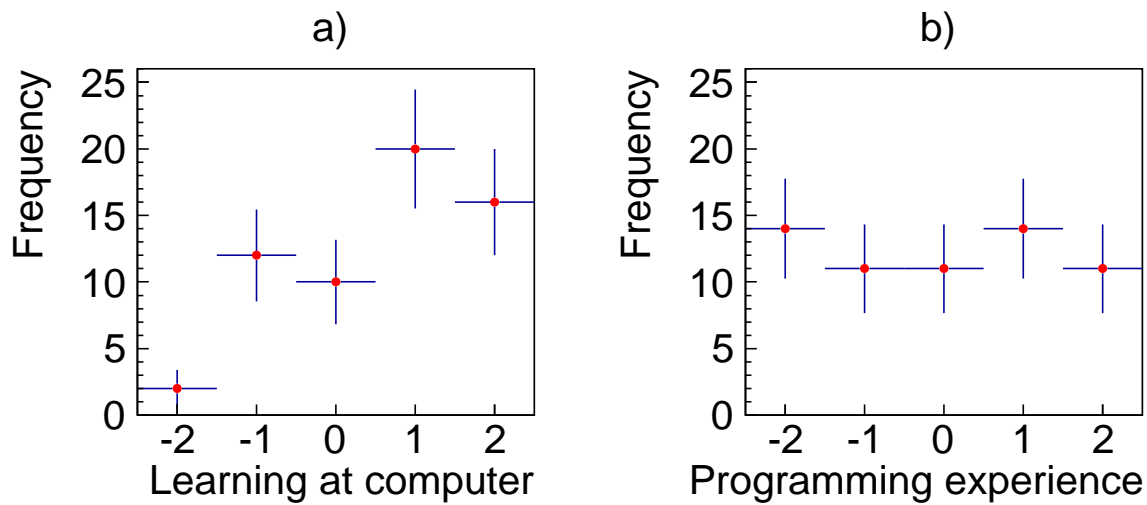

Figure 3: Distributions of students' self-assessment on a) how well they learn at the computer, b) if they have substantial programming experience.

*5.3. Student assessment of the VISPA learning approach*

*5.3.1. Technical Aspects.* The students' experience with the VISPA technology was positive overall. About three quarter of them reported that the VISPA server system was working for them when they needed it, while 5% said they experienced technical problems. 75% of the students could work using the web browser of their own computer, while 16% said they were unable to do so. The students were satisfied with the support (75%). Most of them had sufficient programming experience for performing the exercises (56%).

*5.3.2. Self-assessment of working individually or in teams, and out-of-classroom study time.* In the beginning of the course we had recommended team work in the classical sense of working together at the same location and time, as well as working together virtually through the web system using the students' public folders. About half of the students reported they performed the exercises in team work, the other half preferred working individually on the exercises (figure 4a). A small fraction of students used virtual team work (11%).

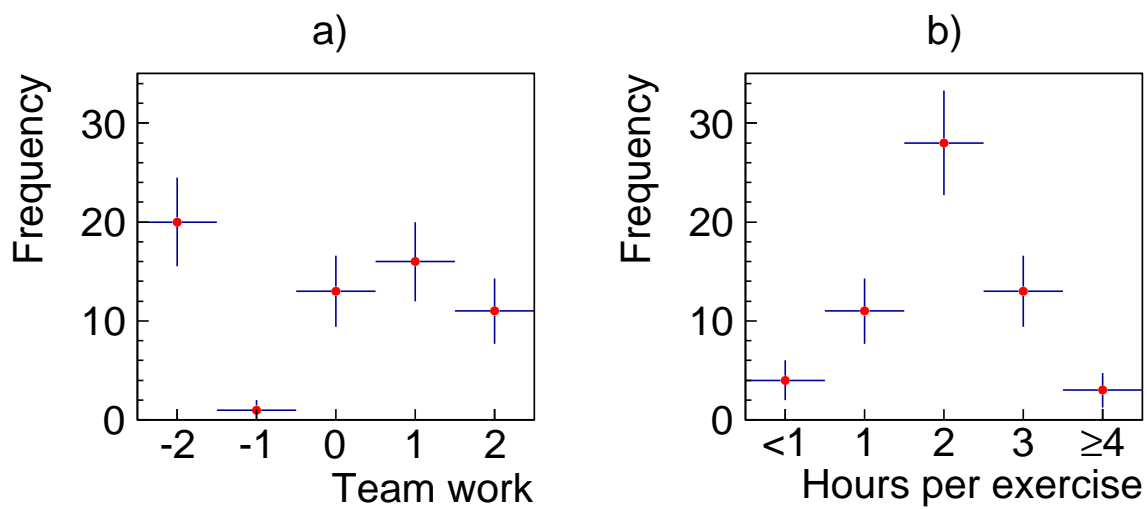

Figure 4: a) Exercises worked on in teams, b) working hours per exercise.

Most of the students used 2 hours per exercise, which we had envisioned as a typical time for a single task. About 25% of the students reported they were faster than 2 hours, and 27% needed more time than the majority (figure 4b).

*5.3.3. Learning benefit and learning style as perceived by the students.* In the students' assessment of their learning benefit in physics, about half of them rated the exercises to deepen the contents of the course lectures (figure 5a). Two-thirds of the students appreciated that they were approaching actual research with the exercises (figure 5b). The latter result was interesting with respect to the physics major curriculum which so far includes actual research in the following semester term only.

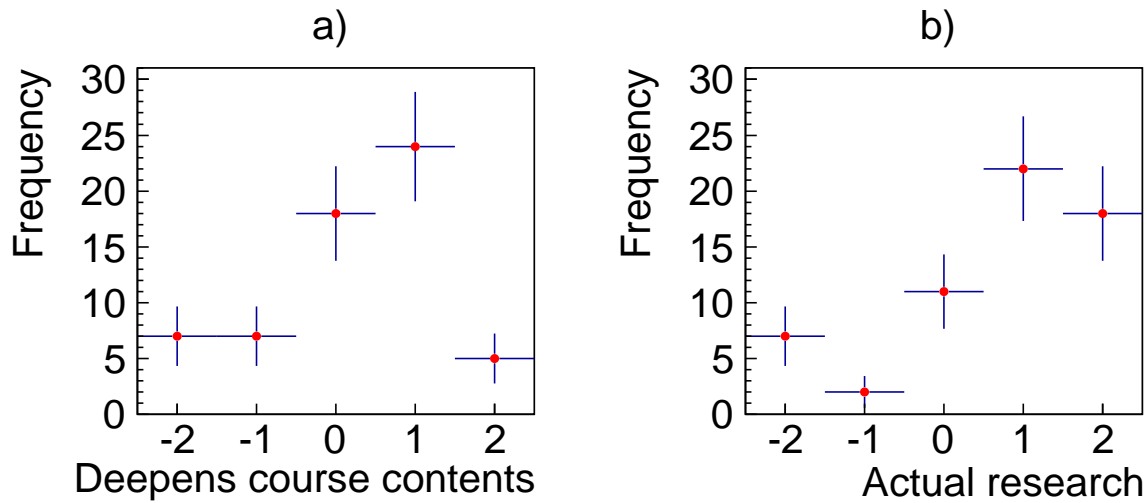

Figure 5: Data analysis exercises with VISPA a) deepen the contents of the physics lectures, b) provide insight into actual research.

When relating the VISPA learning offer with the personal learning style of the student, half of them reported that they enjoyed performing the exercises. One quarter of the students found the exercises to accommodate their individual learning style.

*5.4. Free text comments*

Within the survey, 36 free text comments were received. Out of these comments, we assessed 19 as positive, 11 as negative, and 6 as neutral remarks. Many free comments directly relate to the numerical survey discussed above.

Valuable comments on the workflow were received, e.g. with respect to the number of mouse clicks needed for performing the data analysis exercises (17 comments). The server-client connection was rated being occasionally slow by 5 comments. Two comments mentioned initial technical problems, e.g. during registration. The response of the support team received 4 positive comments.

Concerning the exercises, 9 students called them fun. Four comments judged the exercises as too difficult, and not being suited for every participant. Six participants rated the exercises too easy, or not to contain enough physics (5 comments). Three students wished to receive more information on the physics background of the exercises.

*5.5. Overall rating of the new teaching approach*

The distribution of the overall mark for the data analysis exercises using the VISPA web system is shown in figure 6 on a scale between 1 and 5 with 5 being the best possible mark. An overall positive assessment of the learning project is visible on the right side of the distribution (mark>3). 8% of students gave lowest ratings visible at the left side of the distribution (mark=1). 60% of the students recommended data analysis exercises using VISPA in the course of the following year.

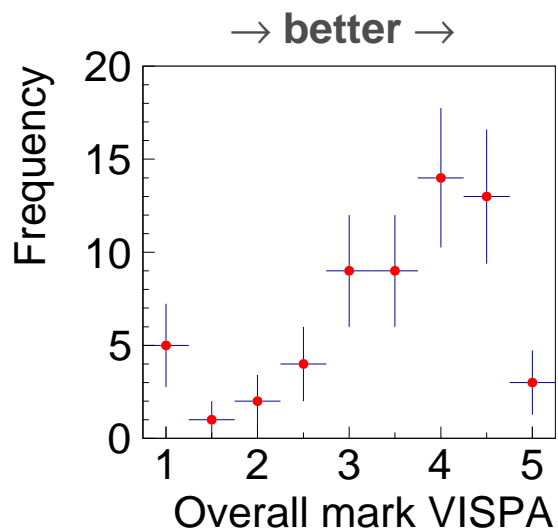

Figure 6: The students' overall rating of the VISPA learning concept on a scale with the highest value being the best mark.

*5.6. Factors influencing students' acceptance of the learning concept*

In order to understand criteria of the students for their overall judgment of the computer exercises, we correlated all questions with the overall mark for the VISPA project and calculated the Pearson's correlation coefficients. Obviously, the recommendation of the students for a future course using VISPA in data analysis exercises was strongly correlated by 85% with the overall mark.

The overall mark for VISPA is also correlated with the questions on the techniques, e.g. whether the VISPA internet platform is suited for data analyses (83%). All questions on the students' learning style are correlated with the overall mark for VISPA. The largest correlation corresponds to the fun aspect, which accounts to 79%.

The students' benefit for their learning of physics plays an important role in the judgment as well. The overall mark for VISPA is correlated, e.g. with their assessment of deepening the contents of the physics lectures by 69% (figure 7a).

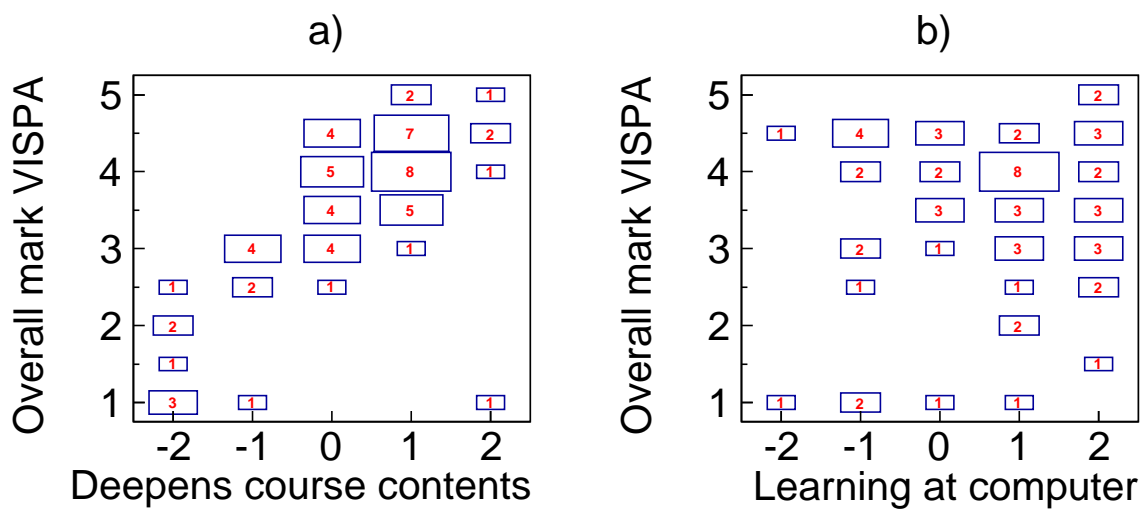

Figure 7: Correlation of the overall mark for the VISPA learning concept a) with the deepening of the contents of the physics lectures, b) with the self-assessment of how well the students learn at the computer.

It is interesting to note that all other correlations are weaker by far than the leading criteria above. The self-assessment on the daily usage of computers and computer programs, the questions on individual versus team work or on the experience with the exercises with respect to working hours all have correlations well below 50%.

However, an interesting observation is presented in the correlation distribution of the overall mark for VISPA with the self-assessment of the learning abilities at the computer shown in figure 7b. A large fraction of students that evaluated themselves as learning well with a computer evaluated the VISPA project positively. Students who reported that they do not learn well with a computer partly liked the project, and a small fraction of them disliked the project (overall mark=1).

*5.7. Influence of programming experience on working style and working hours*

Further interesting aspects resulted from the self-assessment of the students on their programming abilities. Students who brought substantial programming experience had the tendency to work individually (figure 8a). These students also spend less time on the exercises. Students with less programming experience had the tendency to work in teams. For them there is a tendency for spending more time on the exercises as shown in figure 8b.

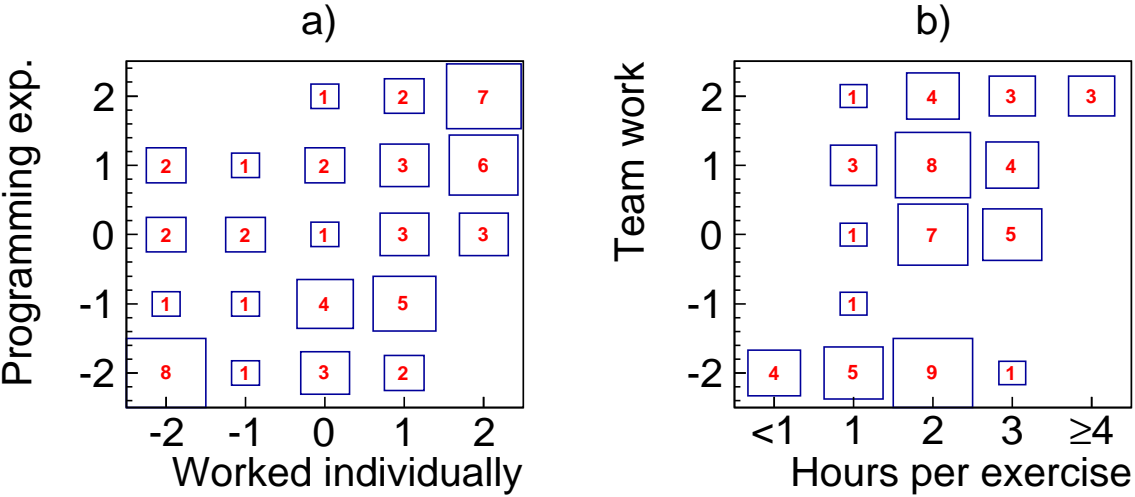

Figure 8: a) Correlations of the self-assessment of the students' programming experience with working individually on the exercises. b) Correlation of working in a team with the working hours per exercise.

# 6. Summary

*6.1. Assessment of the learning project*

Academic teaching shows increasing efforts to include computer-based learning concepts. The learning project reported in this work was designed to complement a $3^{rd}$ year bachelor physics course by data analysis exercises. The new internet platform for developing data analyses provided by the supporting team worked well to successfully perform the educational project.

The majority of the students participating in the exercises and in the survey did the data analyses tasks, and marked the project positively. Their overall mark for the project correlates with the criteria technology, usefulness with respect to their personal learning style, and usefulness with regard to physics learning.

Concerning the offer of an additional learning channel, the evaluation exhibits a majority of students who benefited from the project, and a small fraction of students who did not benefit. This is related to a self-assessment of how well students feel able to learn with a computer.

A further interesting aspect of the learning project was the preferred style of working on the exercises. While students with substantial programming experience preferred to work individually, less experienced students found their way forward through working in teams. In this way, appropriate individual solutions were found for solving the exercises.

It was interesting to note that students appreciated getting involved into actual research through the data analysis exercises for the first time in their university career.

For future courses, we conclude from the survey that many students took advantage from the learning concept to include data analysis exercises. The working hours per exercise were appropriate such that the current level of the exercises should be kept as the base line. However, as the students' programming experience is broadly distributed future iterations of the exercises should facilitate an enhanced flexibility to cover a wide range of students' abilities. As the individual learning styles are broadly distributed, the option of successfully completing the course with text-based exercises exclusively should be retained.

The course of the following year will be evaluated in the same way as described above, and comparisons will be made between this first study and its follow-up course.

*6.2. Aspects concerning the project future*

The development of physics data analysis in a web browser without software installations is a novelty that opens numerous possibilities for realizations of blended learning in experimental physics.

Moreover, communication on data analysis strategies by exchanging a web address offers a new form of exchange among students and between students and faculty. A data analysis can be performed regardless of location and time by the working parties, and the analysis can be evaluated and also developed further.

Even in a larger scientific context one can imagine accompanying publications with web access to the data, the analysis strategies and even the programs. With such analysis preservation, a new form of openness and transparency in physics could be established, which may eventually have an impact on many fields of science and their academic teaching approaches.

## 7. Acknowledgements

The authors wish to thank H.-P. Bretz, M. Komm, and J. Steffens for contributions to the development of the VISPA internet platform, and for fruitful discussions. We are also grateful for financial support to the Ministerium für Wissenschaft und Forschung des Landes Nord-Rhein-Westfalen, the Bundesministerium für Bildung und Forschung (BMBF), the Helmholtz-Allianz Physics at the Terascale, and the Deutsche Forschungsgemeinschaft (DFG).

**References**

Bender H, Sauter A and Sauter W 2002 *Blended Learning. Effiziente Integration von E-Learning und Präsenztraining* (Luchterhand, Neuwied)

Bretz H-P et al. 2012a A Development Environment for Visual Physics Analysis *JINST* **7** T08005

Bretz H-P et al. 2012b A Server-Client-Based Graphical Development Environment for Physics Analyses (VISPA) *J. Phys.: Conf. Ser.* **396** 052015 http://vispa.physik.rwth-aachen.de

Brun R and Rademakers F 1996 ROOT – An Object Oriented Data Analysis Framework *Nucl. Instrum. Meth.* A **398** 81 http://root.cern.ch

Busse S 2012 Lernen im Netz *RWTHinsight 4/2012*. Retrieved Jan 3, 2014, from http://issuu.com/rwth/docs/insight_12_04/8

Johansson K E, Kobel M, Hillebrandt D, Engeln K and Euler M 2007 European Particle Physics Masterclasses make students Scientists for a Day *Phys. Educ.* **42 No 6** 636-644 http://www.physicsmasterclasses.org

Python Programming Language – Official Website 2014 Retrieved Jan 3, 2014, from http://www.python.org

Wen-Yu Lee S et al. 2011 Internet-based Science Learning: A review of journal publications *International Journal of Science Education* **33:14** 1893-1925